\documentclass[twocolumn,pre,superscriptaddress]{revtex4}
\usepackage{graphicx}% Include figure files
\usepackage{dcolumn}% Align table columns on decimal point
\usepackage{bm}% bold math
\usepackage{subfigure}

\newcommand{\figwidth}{8.5cm}
\newcommand{\figsubwidth}{4cm}

\begin{document}

\preprint{APS/123-QED}

\title{Community Detection in Complex Networks Using Genetic Algorithms}

\author{Mursel Tasgin}
\author{Amac Herdagdelen}
\author{Haluk Bingol}%
\affiliation{%
Department of Computer Engineering\\
Bogazici University, Turkey
}

\date{\today}

\begin{abstract}
Community detection is an important research topic in complex 
networks. We present the employment of a 
genetic algorithm to detect communities in complex networks which is based 
on optimizing network modularity. It does not need any prior knowledge about 
the number of communities. Its performance is tested on two real life networks with known community structures and a set of synthetic networks. As the performance measure an information theoretical metric, variation of information, is used. The results are promising and in some cases better than previously reported studies.
\end{abstract}

\maketitle

\section{Introduction}
\label{sec:introduction}
Community structure detection is one of the hot topics that have created a 
great interest in complex network studies. A community is loosely defined as a 
group of vertices with a high density of in-group and a low density of 
out-group edges and their identification in complex networks calls for 
techniques borrowed from physics and computer sciences \cite{Dorogovtsev2002,Clauset2004, Xu2005}.

Different methods and algorithms have been proposed to reveal the underlying 
community structure in complex networks. A crucial part of the algorithms is 
how they define a community \cite{Duch2005}. There are different formal definitions of a 
community \cite{Wasserman2004}. In this paper, we use a quantitative definition proposed by 
Girvan and Newman which makes use of a measure called network modularity~\cite{Newman2004}. The \textit{network 
modularity} $Q$ is defined as
\begin{equation}
Q=\sum_i(e_{ii} -a_i^2) 
\end{equation}
where the index $i$ runs over all communities, $e_{ii}$ is the fraction of edges that connect two nodes within group $i$, while $a_i$ is the 
fraction of edges that have at least one endpoint within the group. Some of 
the recent community detection algorithms like Newman's fast algorithm for 
detecting communities, the algorithm for very large networks, and the 
algorithm using Extremal Optimization use the network modularity as quality 
metric \cite{Clauset2004,Duch2005,Newman2004b}. Calculation of the network modularity is less time 
consuming than the edge betweenness centrality used in Girvan-Newman (GN) 
algorithm~\cite{Newman2004}.

In this paper, we propose a new community detection algorithm which tries to 
optimize the network modularity by employing genetic algorithms. Unlike the 
previous methods, the new algorithm does not require the number of 
communities present in a graph. The number of communities comes as an 
emergent result as the modularity value is optimized.

\section{Background}
\label{sec:Background}
Our algorithm is based on the optimization of network modularity $Q$ by 
employing a genetic algorithm. The algorithm produces a partition on the set of vertices of graph. Success of a community detection algorithm is defined as the closeness of the partition generated by the algorithm to the partition corresponding to the real community structure. Therefore before describing the algorithm, let us introduce the terminology on partitions that we will use throughout this paper. 

\subsection{Representation}
\label{sec:Representation}
We are given an undirected graph $G(V, E)$ with $\left| V \right| = n$ vertices and $\left| E \right| = e$ edges. Assuming a fixed ordering of the vertices $V = \{v_1 ,v_2 , \cdots ,v_n \}$, any partition $\Omega$ of $V$ can be represented by a vector $\bm{\kappa} =\left[ \kappa^1 \: \kappa^2  \: \cdots  \: \kappa^n\right]$ of $n$ dimensions where $\kappa^i \in \{1, 2, \cdots, \left| \Omega \right|\}$ is the index of the cluster of the vertex $v_i$. The vertices $v_i$ and $v_j$ are in the same cluster if and only if $\kappa^i=\kappa^j$. Note that this is not a canonical representation, meaning that there are many different vectors corresponding to the same partition. An example would be helpful: Let $V=\{v_1 ,v_2 ,v_3 ,v_4 \}$ be the set of vertices in the given order and consider the partition $\Omega = \{\{v_1 ,v_3 \},\{v_2 \},\{v_4 \}\}$. Then, although they are different, the vectors 
$\bm{\kappa_1} =\left[1  \: 2  \: 1  \: 3\right]$, 
$\bm{\kappa_2} =\left[4  \: 2  \: 4  \: 3\right]$ and 
$\bm{\kappa_3} =\left[2  \: 1  \: 2  \: 4\right]$
represent the  partition. 

Let $\Gamma$ be the partition of $V$ that corresponds the real community structure of the underlying the graph. An element $ \gamma $ of $\Gamma$ is called a \textit{community}. In total, there are $\left| \Gamma \right|$ elements corresponding to  $\left| \Gamma \right|$ communities. The purpose of the community detection is to build an estimation of the partition $\Gamma$ based on the topology of the graph. Let $\Omega$ be the partition that represents our estimated community structure. An element $ \omega $ of $\Omega$ is called a \textit{cluster}. Note that the number of clusters $\left| \Omega \right|$ does not have to be equal to the number of communities $\left| \Gamma \right|$.

In the most general case, the output of our algorithm will be a vector $\bm{\kappa}$ of length $n$ which corresponds to our estimated partition $\Omega$ of the vertex set. We assume that the number of communities $\left| \Gamma \right|$ is unknown but has to be estimated by the algorithm.

\subsection{Distance Metric for Partitions}

The output of our algorithm is a partition $\Omega$ of the vertices in the graph. The evaluation of a resulting clustering $\Omega$ and a given community structure $\Gamma$ of a network is not straightforward because it is not always clear which cluster corresponds to which community and how to deal with mixed clusters which contain 
members of two or more communities. Even the number of clusters $\left| \Omega \right|$ and the number of communities  $\left| \Gamma \right|$ may differ. 

Two important issues regarding the evaluation of a clustering algorithm is the accuracy and the precision of the 
algorithm. \textit{Accuracy} is a measure of the success of an algorithm in 
clustering the members of the same community together without any separation 
(i.e. intra-cluster scatter). \textit{Precision} is a measure of the success of an 
algorithm in creating homogeneous clusters which contain the members of the 
same communities (i.e. inter-cluster scatter). 

In order to understand these concepts, consider two extreme cases.
The partition of singletons, where each vertex is a cluster by itself, that is $\left| \Omega \right| = n$ , is very precise since no cluster contains elements of more then one communities. On the other hand it is very inaccurate since the elements of any community are scattered into many clusters. The second extreme is the case where the partition is composed of single cluster only,  $\left| \Omega \right| = 1$. All the elements of any community are in the same cluster, that is very imprecise. But all the communities are in the same cluster which means it is very accurate.

%An algorithm which returns assigns each 
%vertex to a different cluster will be very precise (every cluster consists 
%of exactly one community) but also very inaccurate (all members of each community 
%will be placed in a different cluster). At the other extreme, an algorithm 
%which assigns each vertex to the same cluster will be very accurate (all 
%members of each community stay together) but also very imprecise because the 
%only cluster will consist of members of all communities.

\subsubsection{Variation of Information}
We decided to employ an information theoretical metric called variation of information $S$ introduced in Ref.~\onlinecite{Meila2007} specifically oriented to compare results of different clusterings. By using $S$, it is possible to calculate a distance between two partitions. Before proceeding further, let us define $S$ more precisely. Let $\Gamma$ and $\Omega$ be two partitions of the set $V = \left\{ v_1, v_2, \cdots, v_n \right\}$. Let $\Gamma = \left\{ \gamma_1, \gamma_2, \cdots, \gamma_{|\Gamma|} \right\}$ be the set of communities and $\Omega  = \left\{ \omega_1, \omega_2, \cdots, \omega_{|\Omega|} \right\}$ be the set of clusters. Consider the partition $\Omega$ and a randomly picked element $v_i$ from $V$. Without any other information, our uncertainty about which cluster of $\Omega$ the vertex $v_i$ is assigned to is shaped by the distribution of the partition $\Omega$. For example, if all vertices are assigned to the same cluster then there is no uncertainty. If each cluster receives an equal number of vertices (homogeneous distribution) then the uncertainty is at a maximum. To measure the uncertainty, we can use information entropy which is a well known metric of uncertainty. In Ref.~\onlinecite{Meila2007}, \textit{the entropy associated with a partition} $\Omega$ is denoted by $H(\Omega)$ and defined as follows:

\begin{equation}
H(\Omega) = -\sum_{\omega \in \Omega}p(\omega)\log(p(\omega))
\end{equation}
where $p(\omega)=\left| \omega \right| / n$ is the probability that a randomly chosen vertex is assigned to the cluster $\omega$ in partition $\Omega$. The base of the logarithm is irrelevant in our context and we employ the binary logarithm function so the unit of $H$ is bit.

Now, imagine that we have the knowledge about community partition $\Gamma$ of the same graph and we know which community $\gamma$ the randomly picked vertex $v_i$ is assigned to in partition $\Gamma$. This allows us to calculate the \textit{conditional entropy} $H(\Omega|\Gamma)$ defined as

\begin{equation}
H(\Omega|\Gamma) = -\sum_{\omega \in \Omega} 
\sum_{\gamma \in \Gamma}
p(\omega, \gamma)
\log(p(\omega | \gamma))
\end{equation}
which is the amount of entropy (i.e. uncertainty) remaining in $\Omega$ given our knowledge about $\Gamma$. The joint probability $p(\omega, \gamma)$ is the probability that our randomly selected vertex is assigned to cluster $\omega$ in partition $\Omega$ and to cluster $\gamma$ in partition $\Gamma$. The conditional probability $p(\omega|\gamma)$ is the probability that our randomly selected vertex is assigned to cluster $\omega$ in partition $\Omega$ given that we know it is assigned to community $\gamma$ in partition $\Gamma$. The conditional entropy $H(\Omega|\Gamma)$ is always non-negative. It is 0 when the knowledge about $\Gamma$ perfectly determines $\Omega$. $H(\Gamma|\Omega)$ is defined similarly.

The \textit{variation of information} $S$ is defined as
\begin{equation}
S(\Gamma, \Omega) = H(\Gamma|\Omega) + H(\Omega|\Gamma)
\end{equation}

We can use $S$ to combine precision and accuracy values into a single metric. The conditional entropy $H(\Gamma|\Omega)$ is the amount of our uncertainty in $\Gamma$ given our knowledge of $\Omega$. It can be used to measure the precision. If an algorithm assigns each vertex to a different cluster in $\Omega$ then knowing $\Omega$ completely determines $\Gamma$. The conditional entropy $H(\Omega|\Gamma)$, on the other hand, can be used to calculate the accuracy of the algorithm. Lower values of $S$ correspond to less uncertainty (e.g. $S=0$ means the two partitions are identical, hence there is no uncertainity), higher values correspond to more uncertainty. Interested readers may refer to Ref.~\onlinecite{Meila2007} for further discussion of the metric. We should also note that just before the writing of this paper was completed, we came across a recent study which also incorporates $S$ as a metric \cite{Karrer2007}.

\subsection{Genetic Algorithms}

Genetic algorithms (GA), as proposed in Ref.~\onlinecite{Holland1992}, are a set of optimization techniques inspired by the biological evolution. Successful applications have been made to a wide variety of problems including energy minimization~\cite{Morris1996}, traveling salesman problem~\cite{Wu2004,Grefenstette1985}, neural networks and cryptology~\cite{Ruttor2006}, process scheduling~\cite{Wang1995} and other various optimization processes~\cite{Mitchell1998}.
 
In a typical GA, there is an objective function (called as 
the \textit{fitness function}) to be optimized and a set of candidate solutions which are encoded as 
a kind of numerical chromosome. At the start of the algorithm, one begins by 
generating a random population of candidate solutions. The candidates are 
evaluated by using the fitness function. The next generation of candidate 
solutions is generated by applying certain biologically inspired 
manipulations to the current pool of candidates and the solutions with 
higher fitness values have higher chances to be represented in the next 
generation. Here the fitness function plays the role of reproductive fitness 
in Darwinian natural selection. Repeated rounds of fitness evaluation, 
reproduction, and selection cause the initially random population of 
candidate solutions to evolve toward a population enriched in more optimal 
(in terms of fitness function) solutions. The main operations used to 
generate new potential solutions are analogues of \textit{point mutation} (random changes to some 
part of the numerical chromosome) and \textit{crossing over} (forming new chromosomes by combining 
segments of existing ones). Much of the skill in using this approach rests 
in setting up the relationship between the chromosomes and the parameters of 
the optimization problem in such a way that the evolution operations, point 
mutation and crossing over, generate better, or at least not substantially 
poorer, candidate solutions. Genetic algorithms are particularly attractive 
for problems such as combinatorial optimization, where the objective 
function has little or no smooth structure. GAs have the further charm that 
they require no arbitrary convergence criteria. They are not, however, 
parameter-free: one must choose, for example, population sizes, rates of 
mutation, and numbers of generations.

\section{The Algorithm}
It is possible to employ different kinds of genetic algorithms for a particular problem and for every implementation there will be several model parameters like the number of chromosomes, the rate of mutation, or the rate of crossing over as we will explain later. Unfortunately, the values of these parameters are not dictated by the problem at hand but has to be set to some values (somewhat arbitrarily) by us. Since our purpose in this study is to show that genetic algorithms are a viable approach for the community detection algorithm, we will suffice to employ values that are found by trial and error. We will not neither try to study the effect of differing parameter values nor propose a general way to come up with "good" parameter values. This kind of analysis is out of the scope of this paper.

We use $n$-vectors $\bm{\kappa}$ as the chromosomes and the network modularity $Q$ as the fitness function of our genetic algorithm. The \textit{population} $P = \left\{\bm{\kappa_1}, \bm{\kappa_2}, \cdots, \bm{\kappa_p}\right\}$ is the set of all chromosomes. Note that the \textit{population size} $p = \left| P \right|$ is a model parameter. 

\subsection{Initial Population} 
Initially, for all chromosomes, each vertex is put in a different 
cluster. Thus the number of clusters for each chromosome in the 
initial population is $n$. It is a common practice to give the genetic algorithm not a completely random initial starting point but a biased one in order to speed up the convergence. For this purpose, we employed a very simple heuristic. For chromosome  $\bm{\kappa_k}$, we randomly pick a vertex $v_i$ and assign its cluster to all of its neighbors (i.e. $\kappa_k^j \leftarrow \kappa_k^i$ whenever $(v_i,v_j) \in E$). We repeat this operation $\alpha n$ times for each chromosome in the initial population where $\alpha$ is a model parameter and $\alpha=0.4$ is used for the experiments reported in this paper. This operation is extremely fast and results in local small communities. But the resulting clusterings are still far away from being optimal as we will see in the next section.

\subsection{Main Loop}
After the initial population is created, the main loop of the algorithm is repeated $g$ times. Since at each iteration of the loop a new generation is obtained, $g$ is called the \textit{number of generations}:

\begin{enumerate}
\item Apply the fitness function to chromosomes.
\item Sort the chromosomes with respect to the fitness value and take the top $p$. 
\item Save the top $\beta p$ of the the chromosomes for later use.
\item Pair the sorted chromosomes (ie. $\bm{\kappa_k}$ with $\bm{\kappa_{k+1}}$ whenever $k$ is odd, assuming $p$ is even) and apply crossover operation to the pairs.
\item Apply mutation.
\item Combine newly obtain $p$ chromosomes and the previously saved $\beta p$.
\end{enumerate}

Note that the last step is an elitist approach and ensures that the fitness scores of the top $\beta p$ of the child generation will be at least as good as the 
parent population. Note also that except the initial generation, the second step starts with $(1+\beta)p$ chromosomes and ends with $p$ chromosomes. Here $\beta$ and $g$ are model parameters. We used $\beta =0.1$ in this work. Now, let us focus on the crossing over and mutation operators in detail.

\subsection{Crossing over} 
Traditionally, crossover operator takes two chromosomes, merges 
them together and returns two new chromosomes. A crossing over point in each of the chromosomes is selected, and all the elements of the chromosomes after that selection point are exchanged between the two chromosomes.

Unfortunately, in our settings, the encoding of the chromosomes does not 
allow such a straightforward crossing over operation. For each chromosome, 
the clusters to which the vertices are assigned are represented by arbitrary 
integers and the values in two different chromosomes are not compatible as discussed in Section~\ref{sec:Representation}.

Instead of employing a crossing over operation based on mutual exchange, we decided to introduce a \textit{one-way crossing over} operation. One of the chromosomes in the selected pair is called the \textit{source chromosome} $\bm{\kappa_{src}}$ and the other is called the \textit{destination chromosome} $\bm{\kappa_{dest}}$. The crossing over procedure is defined as follows. We pick vertex $v_i$ at random, determine its cluster (i.e. $\kappa_{src}^i$) in the source chromosome and make sure that all the vertices in this cluster of the source chromosome are also assigned to the same cluster in the destination chromosome (i.e. $\kappa_{dest}^k \leftarrow \kappa_{src}^i$, $\forall k \in \left\{k \;|\; \kappa_{src}^k = \kappa_{src}^i \right\}$).

%\begin{enumerate}
%\item Randomly pick a vertex ($v_x$) in the source. and read its cluster Id (say the cluster Id is $i$).
%\item Determine the set of all of the vertices assigned to the same cluster with the picked vertex in the source chromosome. Call this set $S$. (i.e. $S=\left\{ v_y |\omega_{src}^y = \omega_{src}^x\right\}$)
%\item Set the assigned cluster Ids of all vertices in set $S$ to $i$ in the destination chromosome. (i.e. $\omega_{dest}^z \leftarrow \omega_{src}^z, \forall z \in S$)
%\end{enumerate}

The crossing over procedure is repeated $\eta n$ times on the chromosome. The \textit{crossing over rate} $\eta$ is a model parameter and set to $\eta = 0.2$ for the experiments reported in this paper. An example of a crossing over application is given in Table~\ref{tab:cross}. Note that as a result of crossing over, $ v_7 $ becomes in the same community with $ v_4 $.

\begin{table}[htbp]
\centering
\caption{One-way crossing over when $v_4$ is selected}
\begin{tabular}{|c|c|c|c|c|c|c|}
\hline \textbf{v} & & \mbox{$\bm{\kappa_{src}}$} & & \mbox{$\bm{\kappa_{dest}}$} (before) & & \mbox{$\bm{\kappa_{dest}}$} (after) \\ 
\hline 1 & & \textcircled{7} & $\rightarrow$ & 2 & $\rightarrow$ & \textcircled{7} \\ 
\hline 2 & & \textcircled{7} & $\rightarrow$ & 2 & $\rightarrow$ & \textcircled{7} \\ 
\hline 3 & & 2 & & 5 & & 5 \\ 
\hline 4 & $\rightarrow$ & \textcircled{7} & $\rightarrow$ & 8 & $\rightarrow$ & \textcircled{7} \\ 
\hline 5 & & \textcircled{7} & $\rightarrow$ & 3 & $\rightarrow$ & \textcircled{7} \\ 
\hline 6 & & 3 & & 3 & & 3 \\ 
\hline 7 & & 9 & & 7 & & 7 \\ 
\hline 8 & & 9 & & 4 & & 4 \\ 
\hline
\end{tabular}
\label{tab:cross}
\end{table}

\subsection{Mutation} 
We employ a point mutation operator defined as follows: We randomly pick a chromosome $\bm{\kappa}$ to be mutated. Then we pick two vertices $v_i$ and $v_j$ randomly. The cluster of $v_j$ is set to the cluster of $v_i$ (i.e. $\kappa^j \leftarrow \kappa^i$). The mutation procedure is repeated $\zeta n$ times where the \textit{mutation rate} $\zeta$ is a model parameter and set to $\zeta = 0.5$ for the experiments reported in this paper.

\section{Experimental Results}
Here we set several model parameters to seemingly arbitrary values such as the population size $p$. We would like to stress that our aim in this paper is not to fine tune the genetic algorithm but to provide a proof of concept that genetic algorithms are capable of producing compatible results with the previous algorithms. We evaluated our algorithm on two well known datasets and a set of computer generated networks with known community structures.

\subsection{Zachary Karate Club}

The Zachary Karate Club network, which is one of the few data sets with known community structure, is analyzed first in Ref.~\onlinecite{Zachary1977}. The network 
consists of 34 vertices and 78 edges. We ran our algorithm on this dataset 
for 50 times for $g = 250$ generations with population size is set to $p = 100$.

Although our algorithm does not know the number of communities, in all of the runs, the resulting partition $\Omega$ consisted of 2 clusters as it should ideally be. In 49 runs, the clusters perfectly matched the real communities. In one run we observed that one vertex is misplaced.

\subsection{College Football Network}
College football network is built by using the college football matches in 
USA, for Division I during the year 2000 \cite{Girvan2002}. The vertices in the network are 
the college football teams and there is an edge between two teams if they 
played a match during the season. The real community structure is the 
conferences that each team belongs to. The teams tend to play more matches 
with teams that are in the same conference and play less inter-conference 
matches. 

\begin{figure}[thbp]
\centerline{\includegraphics[width=\figwidth]{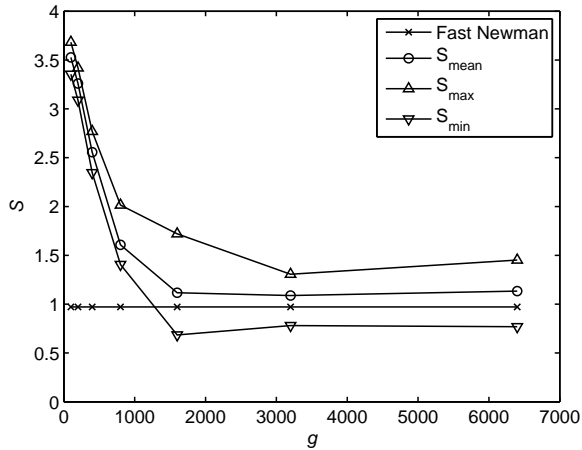}}
\caption{$S$ versus $g$ for college football dataset.}
\label{fig1}
\end{figure}

The dataset consists of 115 vertices with 12 communities (i.e. conferences). We ran the algorithm on this network 10 times with $p=200$ and $g \in \left\{100,200,400,800,1600,3200,6400\right\}$. The $S$ scores obtained are presented with those of the fast algorithm for community detection (Fast Newman) in Fig.~\ref{fig1}. The score $S_{mean}$ is the mean of the $S$ scores we obtained from the 10 runs for each $g$ value. Similarly, $S_{max}$ and $S_{min}$ are the maximum and minimum of these scores respectively.

\subsection{Synthetic Networks}
\label{sec:synthetic_networks}
It is not easy to find large datasets with known community structures. To 
evaluate our algorithm further, we created a set of synthetic networks with 
known community structures. The networks are based on a very simple network 
generation model used in Ref.~\onlinecite{Newman2004b}. They consist of either 128 or 512 vertices. The average degree is set to 16 and the vertices are assigned to 4 predetermined communities of equal sizes. There is a single parameter called $z_{out}$ which regulates the average number of edges that a vertex makes with members of other communities. If $z_{out} =0$ then every vertex has only edges connecting to members of its own community. As we increase $z_{out}$, we obtain networks with weaker community structures. When $z_{out}=12$, the community structure in the topology is completely lost and the network becomes a random network.

We would like to see the performance of our algorithm as a function of $z_{out}$. For this purpose, we set the number of vertices $n=128$ and created 100 networks for each value of $z_{out} \in \left\{ 1, \cdots, 12 \right\}$, that is 1200 networks in total.

\begin{figure}[htbp]
\centering
\subfigure[] % caption for subfigure a
{
    \label{fig2a}
    \includegraphics[width=\figsubwidth]{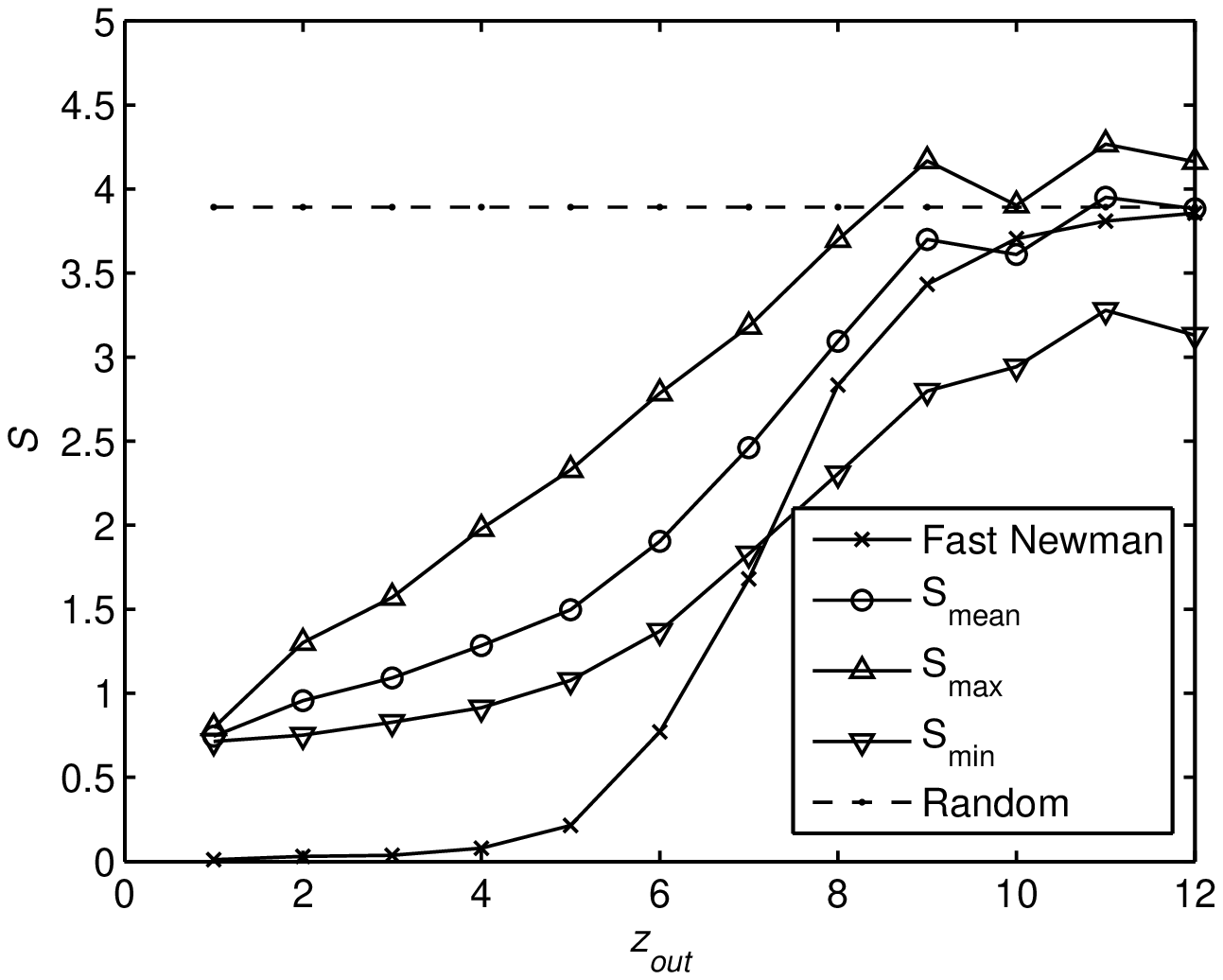}
}
\subfigure[] % caption for subfigure b
{
    \label{fig2b}
    \includegraphics[width=\figsubwidth]{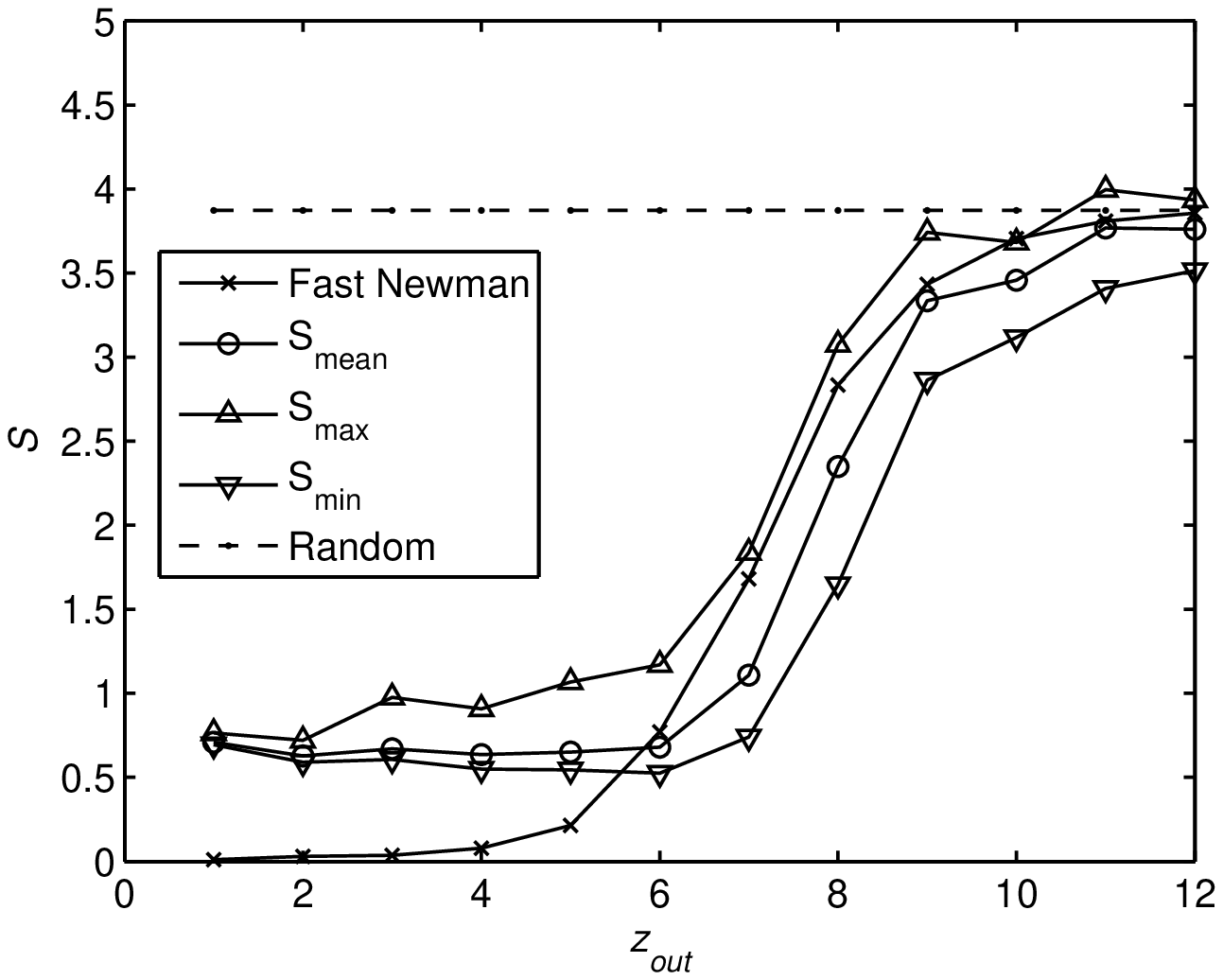}
}
\caption{$S$ versus $z_{out}$ for synthetic networks with $n=128$ and (a) $g=750$, (b) $g=3000$}
\label{fig2} 
\end{figure}

We set population size $p=200$ and the number of generation $g=750$. We ran our algorithm 10 times on each network, calculate $S$ of each run and obtain 10 $S$ values.  We take the minimal $S_{min}$, the mean $S_{mean}$ and the maximal $S_{max}$ values of the 10. Since we have 100 networks for the same $z_{out}$ value, we take the average of $S_{min}$, $S_{mean}$ and $S_{max}$ over 100 networks for a particular $z_{out}$.

The resulting scores are presented in Fig.~\ref{fig2a} as a function of $z_{out}$. The scores obtained by Fast Newman are also given for comparison \cite{Newman2004b}. We also include the scores of a dummy algorithm which assigns each vertex randomly to one of the four clusters.  

The small gap between the maximal and minimal $S$ scores suggests that the performance of our algorithm is robust and does not change from one run to another significantly. The scores of the genetic algorithm and Fast Newman increase when we increase $z_{out}$ as expected because of the weakening community structure. Note that $S_{mean}$ scores of the genetic algorithm and the Fast Newman's scores are compatible for high values of $z_{out}$ but Fast Newman significantly outperforms the genetic algorithm for lower values of $z_{out}$. It seems that when the underlying community structure is strong (i.e. the problem at hand is trivial) the genetic algorithm is unable to converge to a solution as optimal as Fast Newman can find. When the community structure is weakened, the difference between the two algorithms disappears. The question whether this behavior is due to a lack of the fine tuning of the model parameters or an intrinsic property of our genetic algorithm calls for further investigation.

%\begin{figure}[htbp]
%\centerline{\includegraphics[width=\figwidth]{fig2.eps}}
%\caption{$S$ versus $z_{out}$, $n=128$ and $g=1500$ generations. The dashed line is the score obtained by a random clustering algorithm with 4 clusters.}
%\label{fig2}
%\end{figure}

%??
%These 10 $S$ values are used to calculate three different scores for each network:
%obtained for the network during the 10 runs. 
%Then we calculated the average of each of the three scores over each batch of networks. 
%
%
%\begin{figure}[htbp]
%\centerline{\includegraphics[width=\figwidth]{fig3.eps}}
%\caption{$S$ versus $z_{out}$, $n=128$ and $g=3000$ generations. The dashed line is the score obtained by a random clustering algorithm with 4 clusters.}
%\label{fig3}
%\end{figure}

What is the response of our algorithm to increasing the number of generations? In order to give an idea we present Fig.~\ref{fig2b} which contains results obtained in the same way but this time with $g=3000$ generations for each run. The results are qualitatively similar but the $S$ values of the genetic algorithm is lower in general. The improvement in the scores of our algorithm suggests that it is possible to obtain better solutions by increasing the number of generations.

\begin{figure}[htbp]
\centerline{\includegraphics[width=\figwidth]{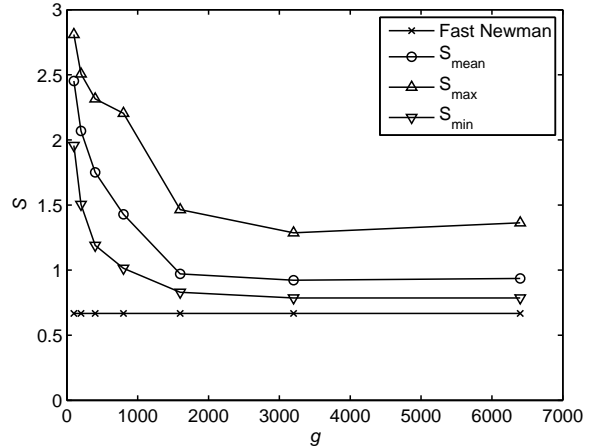}}
\caption{$S$ versus $g$ for synthetic networks with $n=128$, $z_{out} =6$}
\label{fig3}
\end{figure}

In order to analyze this point, we set $z_{out}=6$ and calculated the scores for differing number of generations. In Fig.~\ref{fig3}, we see that the solutions converge after a certain number of generations. The score of the Newman's algorithm is also given by the flat line since it does not depend on our model parameter $g$.

\begin{figure}[htbp]
\centerline{\includegraphics[width=\figwidth]{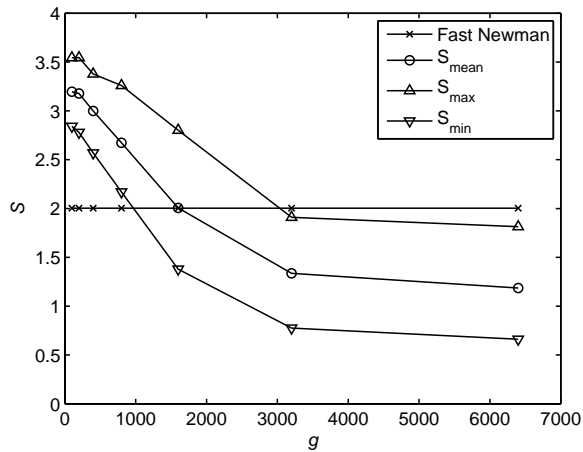}}
\caption{$S$ versus $g$ for synthetic networks with $n=512$, $z_{out} =6$}
\label{fig4}
\end{figure}

To examine our model with larger networks we repeated the same set of experiments with different number of generations ($z_{out} =6$) but this time on networks with $n=512$ vertices. In Fig.~\ref{fig4}, we see that our algorithm still provides results comparable with (and even better than) Fast Newman.

Note that, for fast algorithm of Newman, we use our knowledge on number of real communities by cutting the dendogram just at the right place while the genetic algorithm lacks this information and is still comparable with the fast algorithm.

\section{Conclusions}
In this paper, we proposed a new community detection algorithm, which 
tries to optimize network modularity using genetic algorithm methods. The 
contribution of this study is the introduction of a genetic algorithm for 
the community detection problem which does not require any information about 
the number of communities in the network. The results are compatible with 
previously introduced methods. Thus the employment of genetic algorithms for community detection problem is a viable approach.

\begin{acknowledgments}
This work was partially supported by Bogazici University Research Projects 
under the grant number 07A105 and was partially based on the work performed in the framework of the FP6 project SEE-GRID-2, which is funded by the European Community (under contract number INFSO-RI-031775). The authors also thank to anonymous referees for their enriching comments and suggestions.
\end{acknowledgments}

\bibliography{tasgin2007}

\end{document}